# Optomechanical Design of the DragonCam Microscopic Camera

M. J. Clark*[a], M. A. Ravine [a], M. A. Caplinger [a], B. A. Lindenfeld[b], J. D. Laramee[b], R. S. Bronson[c], A. D. Giglio[c], and B. G. Crowther[d]

[a]Malin Space Science Systems, 15253 Avenue of Science, San Diego, CA USA 92128; [b]Motiv Space Systems, 350 N Halstead St, Pasadena, CA USA 91107; [c]Collins Aerospace, 2752 Loker Ave W, Carlsbad, CA USA 92010; [d]Synopsis Optical Solutions Group, 199 S. Los Robles Avenue, Suite 400, Pasadena, CA 91101

## ABSTRACT

The DragonCam Microscopic Camera is an instrument being developed for NASA's Dragonfly mission [1] to Saturn's moon Titan. The Microscopic Camera will be body-fixed to the Dragonfly vehicle and will image the surface at a distance of about one meter (98.6 cm nominal) with a pixel scale of better than 60 microns/pixel and a nominal 52 degree angle to the Titan surface. With the 4.8 µm pixel pitch of the sensor, this is a focal length of about 77.5 mm.

To accommodate range variations due to vehicle pose and surface topography, the Microscopic Camera has a focus mechanism to give it a depth of field (DOF) of about 130mm. Since the Microscopic Camera's boresight is tilted by 52° off the vertical, the optical configuration has a compensating tilted focal plane, taking advantage of the Scheimpflug imaging principle.

The optics are all-refractive with nine elements, a six-element stationary group and a three-element moving group. A plano-plano window seals the optics from the environment and also serves as the substrate for a bandpass filter.

The optomechanical system is derived from the Mars Hand Lens Imager [11]; the moving group is mounted to a linear slide which is translated via a cam follower by the rotation of a cam driven by a stepper motor.

The Microscopic Camera is designed to survive at temperatures as low as -130° C without power. The camera is enclosed in a cavity in the foam insulation covering the spacecraft and looking through a single-pane window. Prior to imaging, the camera will be heated to operating temperature (nominal -30° C) for proper actuation of the mechanism. STOP analysis has been performed to demonstrate that optical performance is maintained after heating.

Software focus merging will be performed in the onboard camera control electronics to minimize image data downlink requirements.

## 1. INTRODUCTION

The DragonCam Microscopic Camera is part of the DragonCam payload currently being developed for NASA's Dragonfly Mission to Saturn's moon Titan. Dragonfly is a nuclear-powered robotic rotorcraft lander managed and built by the Johns Hopkins Applied Physics Laboratory for launch in 2028 for an arrival at Titan in 2034. Upon arrival, Dragonfly will perform controlled flights each Titan day for a period of 3.3 years, eventually traveling several hundred kilometers from its initial landing site. Sampling at different locations, the rotorcraft will investigate the surface chemistry, potentially unlocking clues about the chemical processes that led to the formation of life on planet Earth, using four payloads. These payloads include the Dragonfly Mass Spectrometer (DraMS), the Dragonfly Gamma-Ray and Neutron Spectrometer (DraGNS), the Dragonfly Geophysics and Meteorology Package (DraGMet), and a Camera System being built by Malin Space Science System known as DragonCam .

The DragonCam payload consists of seven cameras connected to a Digital Video Recorder (DVR), all components of MSSS's ECAM modular space camera system. The ECAM product line uses radiation-tolerant, high-reliability components suitable for NASA spacecraft, but offered on a commercial basis with little to no development cost. Multiple types of cameras exist under the ECAM platform, including rolling [3] and global shutter [4] visible cameras, VGA [5] and SXGA [6] Long Wave Infrared (LWIR) cameras, and Short Wave Infrared (SWIR) Cameras, with each camera type having multiple existing optical designs with different Effective Focal Lengths and F-numbers. While notably different from each other, the cameras all share the same high-reliability design methodology, as well as having the same Spacewire electrical interface. This common interface allows the cameras to be controlled by a single ECAM DVR [7] unit, which can interface with up to 8 cameras. The DVR provides and Spacewire, RS-422, or other custom LVDS interface to the spacecraft and operates off unregulated 28V power. In turn, the DVR provides the required data interface to the cameras, converts spacecraft power to regulated 5V power for the cameras, provides up to 500GB of storage, compression algorithms, or other post-processing capabilities. The design reliability, fidelity, heritage, and adaptability of the ECAM system makes it a cost-effective option for the Dragonfly mission.

The DragonCam payload consists of five ECAM-P50 cameras with 5.8mm lens (Down, Side, and Forward WAC cameras), one ECAM-P50 with 12.6mm lens (Forward MAC), and an ECAM-L50 with custom focus mechanism (Microscopic Camera). These cameras are connected to an ECAM-DVR8P, which resides inside the spacecraft. Redundant Navigation Cameras (NavCams) are also being manufactured and delivered by MSSS for performing Terrain Relative Navigation (TRN), but interface directly to the spacecraft.

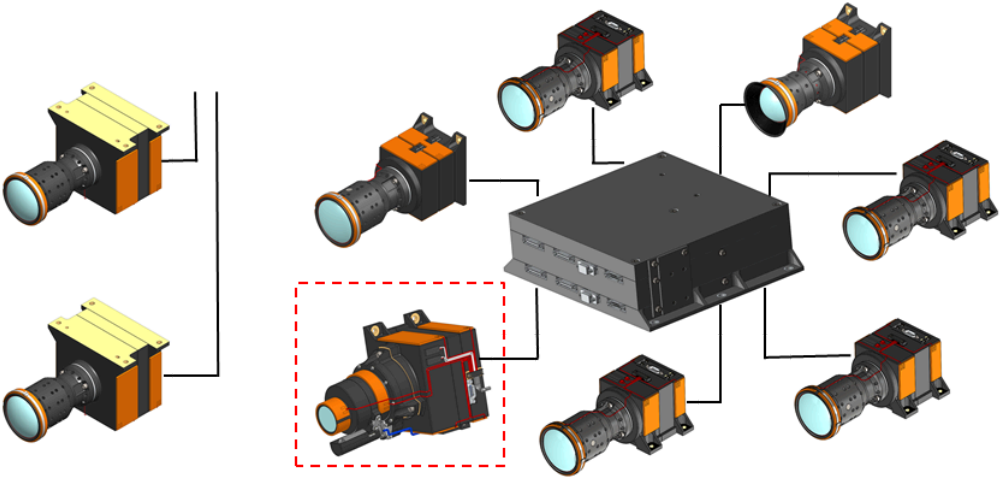

*Figure 1. Cameras being developed by Malin Space Science Systems for the Dragonfly spacecraft. The left-most cameras are the Navigation Cameras, the camera within the red box is the Microscopic Camera, and the remaining cameras are the Forward, Side, and Down Cameras.*

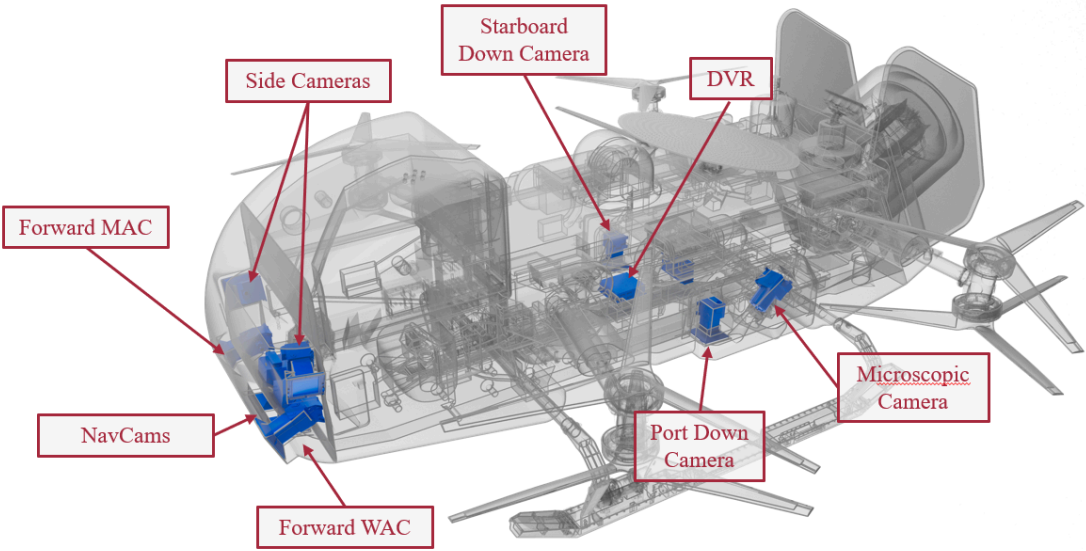

*Figure 2. Locations of camera components within the Dragonfly spacecraft.*

While heritage designs existed for all elements of the DragonCam payload, several customizations were necessary in order to survive and operate in the Titan environment. Much of this customization is owed to the Titan thermal environment and the spectral range of interest on the Titan surface, and specifically for the Microscopic camera, a custom optical design and optomechanical implementation.

The Microscopic Camera is mounted to the side of the Dragonfly rotorcraft and images the drill site of the Sample Acquisition Drill. The camera is mounted approximately 1m from the drill site at a 52 degree angle with respect to the Titan surface normal. An LED array (designed and built by APL) is coaligned with the drill site to provide spectral imaging capability, including UV fluorescence. This allows the Microscopic Camera to capture images at incremental spectral resolutions at night, while taking broadband color images during the day.

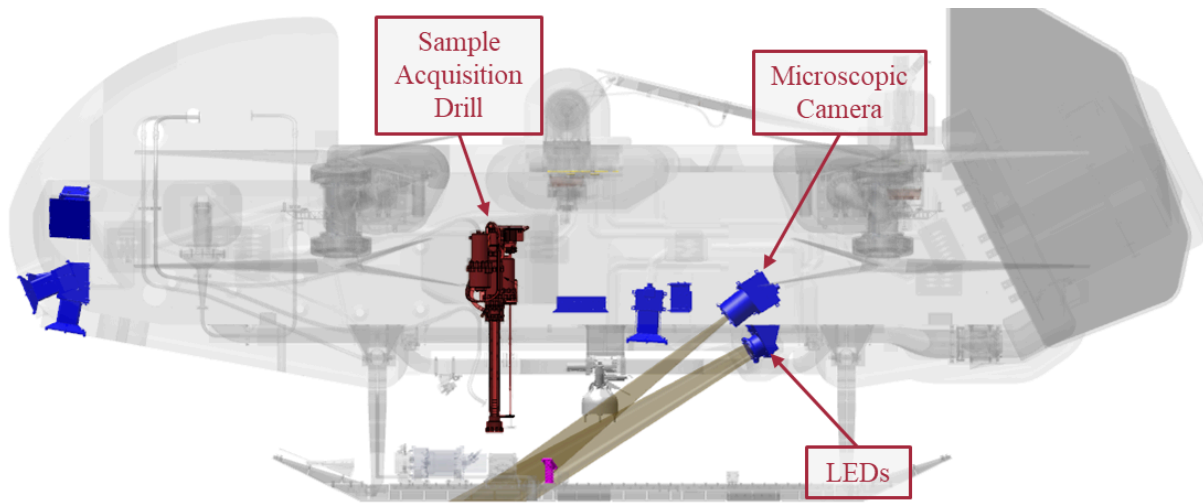

*Figure 3. Side View of the Dragonfly Spacecraft showing the co-locating of the Microscopic Camera, LED array, and Sample Acquisition Drill.*

The Microscopic Camera must survive and operate in harsh conditions. The camera must survive launch loads including random vibration, pyroshock, and quasi-static loads. It must also survive not only the radiation environment encountered during a 6-year cruise in deep space, but also bombardment from neutrons emitted by the spacecraft's Multi-Mission Radioisotope Thermoelectric Generator (RTG). During terrestrial flight on Titan, the unit must survive the sine vibration endurance loading associated with the rotorcraft flight without fatigue. And finally, the camera must be able to survive and operate in a range of thermal environments. These environments range from cruise through the vacuum of space, where temperatures up to +70°C may be encountered, to terrestrial flight in Titan's 1.5bar, -179°C atmosphere [11].

The Dragonfly spacecraft has a warm electronics box that uses waste heat from the MMRTG to heat the vehicle body, while thick insulation minimizes heat loss to the cold Titan environment [12]. The cameras, however, are thermally isolated from this benign environment as they inherently need to be mounted externally to image the scene. To minimize this heat leak, the spacecraft provides a thermally isolated enclosure with a window that the cameras image through. This means that the camera needs to provide enough heater power to warm up and maintain its operational temperature. While each camera has adequate heater power, those heaters cannot be powered on while the spacecraft is hibernating and trickle-charging its batteries. Therefore, the cameras must survive off the heat leak of the spacecraft through the thermal isolators during periods of hibernation. With nothing but a convectively-cooled 3mm thick window between the camera and the -179°C atmosphere, the cameras must survive temperatures as low as -130°C.

The Microscopic Camera will be used for a number of scientific purposes:

- It will acquire color images at 120μm resolution covering an area greater than 5x5cm around each sampling location under UV illumination at night to identify aromatic organics (particularly Polycyclic Aromatic Hydrocarbons, or PAHs) that fluoresce when stimulated by near-UV light.

- It will acquire color microscopic images at 120μm resolution around each sampling location, before and after sample acquisition, similar to the MAHLI instrument on Mars Science Laboratory [10].

- In conjunction with the port Side and Down Cameras, will observe controlled active saltation experiments to determine the sand saltation threshold freestream speed to 0.1m/s (i.e., will observe movement of individual sand grains that are affected by the rotor downwash).
- Will provide tactical imaging used in support of the DrACO payload

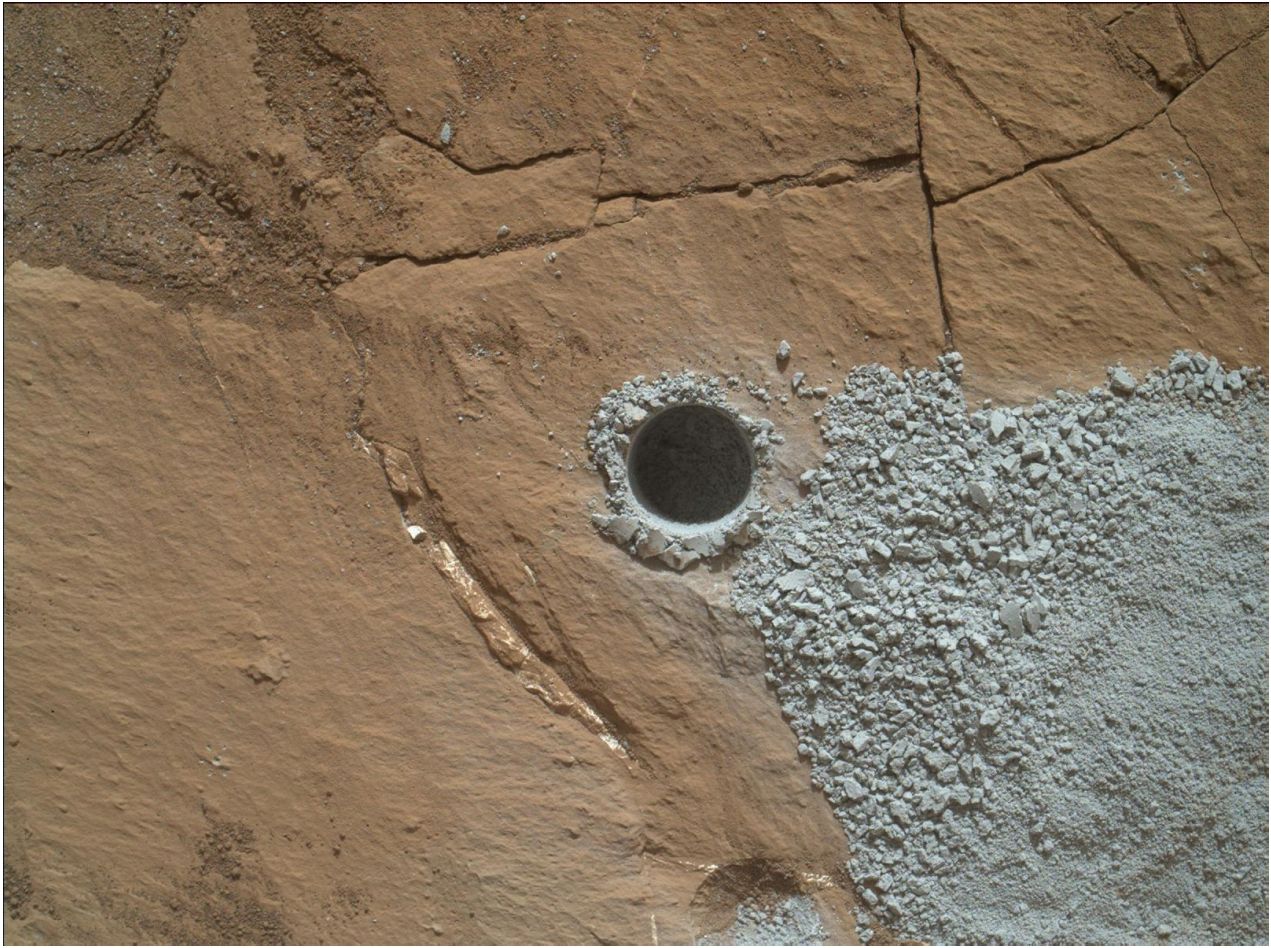

*Figure 4. This MAHLI image shows a drill hole near "Marias Pass" on lower Mount Sharp. Image Credit: NASA/JPL-Caltech/MSSS*

## 2. Dragoncam Microscopic camera design

The Microscopic camera consists of an electrical box that houses three printed circuit boards and a lens assembly that contains the stationary and moving optical groups, motor, and focus mechanism.

The electrical box supports three printed circuit boards and is made of aluminum. The thickness of the enclosure is sized to provide radiation shielding for the electronics without adding unnecessary mass, as the Dragonfly mission was mass-constrained from the onset. The electronics box provides the mounting interface and alignment features to the lens assembly and to the spacecraft. Heaters are affixed to the exterior of the electronics box to warm up the unit from -85°C survival temperature during hibernation, to greater than -30°C operating temperature within 60 minutes. The electronics box also provides the slanted interface to the sensor required for the Scheimpflug optical design.

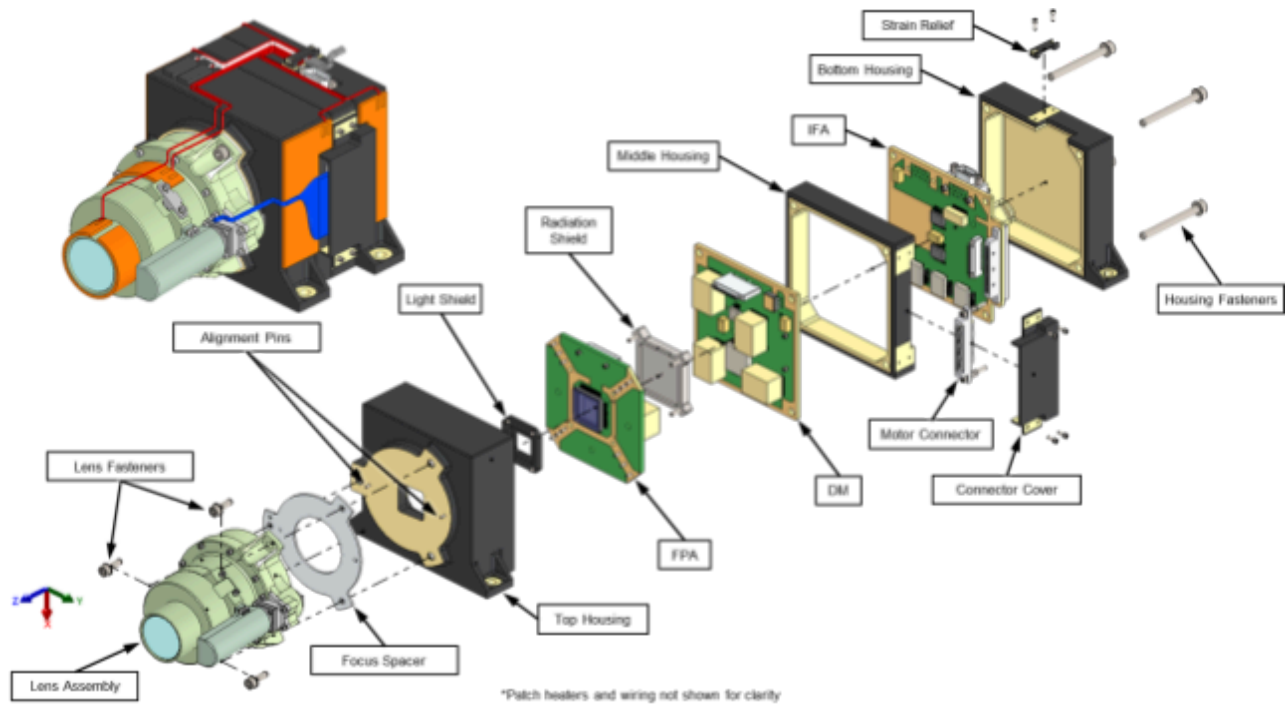

*Figure 5. Exploded view of the DragonCam Microscopic Camera*

Table 1. Design Specifications for the Microscopic Camera

| Parameter | Design Value |
|---|---|
| Pixel Format (pixels) | 2048 x 2048 |
| Pixel Pitch (µm) | 4.8 |
| Effective Focal Length (mm) | 77.5 |
| Focal Ratio | 5.6 |
| Pixel Scale | <60µm/pixel at 986mm |
| Spectral Range (nm) | 440 to 975 |
| Mass (g) | 1040 |
| Instrument Volume (mm) | 105 x 155 x 90 |
| Depth of Field (mm) | 986 ± 65 |
| Angle to Titan Surface Normal (deg) | 52 |
| Optical Design | Scheimpflug |
| Operating Temperature Range (°C) | -35°C to +55°C |
| Survival Temperature Range (°C) | -105°C to +70°C |
| Random Vibration (Launch, g RMS) | 8.11 |
| Shock (G, Q=10) | 1386 @ 2000Hz |

**Mechanical**

The Microscopic camera leverages a heritage focus mechanism design used for five instruments currently operating on the surface of Mars. On the Mars Science Laboratory (Curiosity) Rover, there are three copies, both MastCams and the Mars Hand Lens Imager (MAHLI). On the Mars 2020 (Perseverance) Rover, there are two copies, one in the WATSON instrument and one as part of the SHERLOC instrument (known as the Focus/Cover Mechanism.

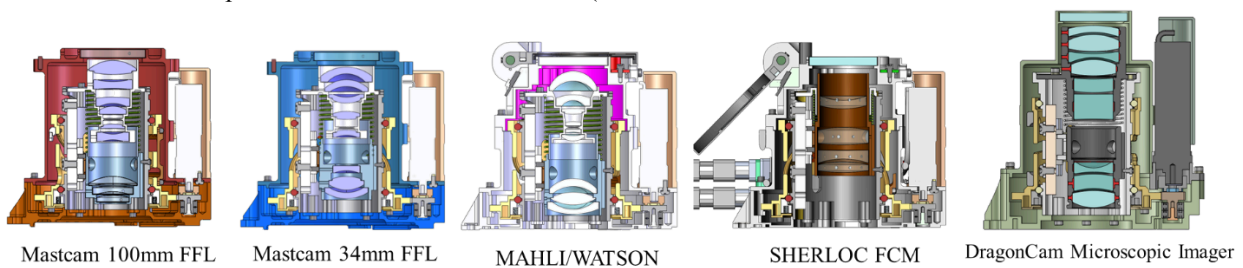

Mastcam 100mm FFL    Mastcam 34mm FFL    MAHLI/WATSON    SHERLOC FCM    DragonCam Microscopic Imager

*Figure 6. Cross-sections of heritage MSSS focus mechanisms.*

Each of these instruments used a different optical design and the Dragonfly Microscopic Camera is no different. Like its predecessors, the Microscopic Camera has three optical groups: a stationary or fixed group, a moving or focus group, and a window that seals the mechanism and serves as a substrate for the bandpass filter. The moving group slides on a linear rail to provide a means of focus adjustment for the instrument. The focus group position is adjusted be actuating a stepper motor. The mechanism transforms the motor rotation into linear movement through means of a Cam-based system.

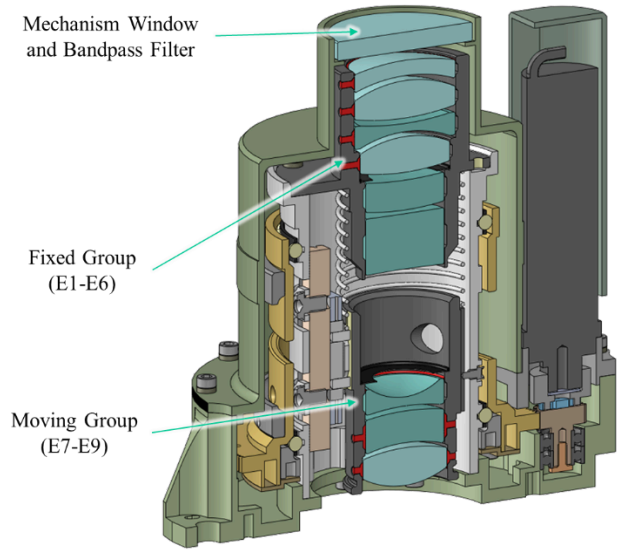
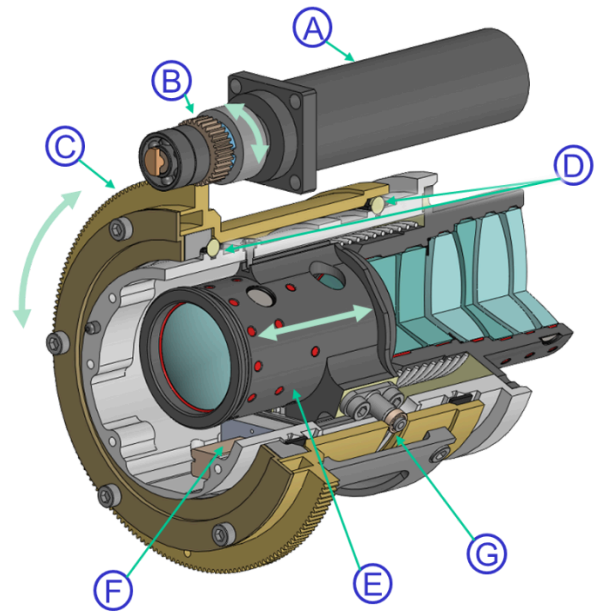

*Figure 7. Cutaway of the Microscopic Camera focus mechanism showing the three optical groups.*

*Figure 8. Diagram illustrating translation of rotational motor output to translational movement of focus group by means of (a) Stepper Motor, (b) Pinion Gear, (c) integral ring gear on cam barrel, (d) integral bearing between cam barrel and inner optical structure, (e) focus group, (f) linear rail, (g) cam follower pin.*

The mechanism is driven by a ½" stepper motor with a 56.25:1 planetary gearhead speed reduction, identical to the motor flown on the Psyche Multispectral Imager Filter Wheel [9] but with increased gear ratio. The Gearmotor is coupled to a pinion gear via a custom Oldham coupler that allows for translational misalignment of rotational axes. The pinion gear is supported by a pair of duplex ground back-to-back radial bearings to alleviate radial loading of the gearmotor output bearings.

The Gearmotor pinion gear meshes with a ring gear that is integral to the Cam Barrel. Backlash of the gear mesh is controlled with mechanical tolerances at the piece-part level. Given the volumetric constraints of the mechanism and the keepout area occupied by the optics, an integral bearing is implemented between the cam barrel and inner optical structure. The integral bearing pair consists of 3-point contact and preloaded 4-point contact bearing arrangements to support the cam barrel. The preload is accomplished with 8 helical springs and eliminates radial/axial movement of the cam barrel during operation and axially constrains the cam barrel, whereas the axially-free 3-point arrangement prevents over-constraint across temperature.

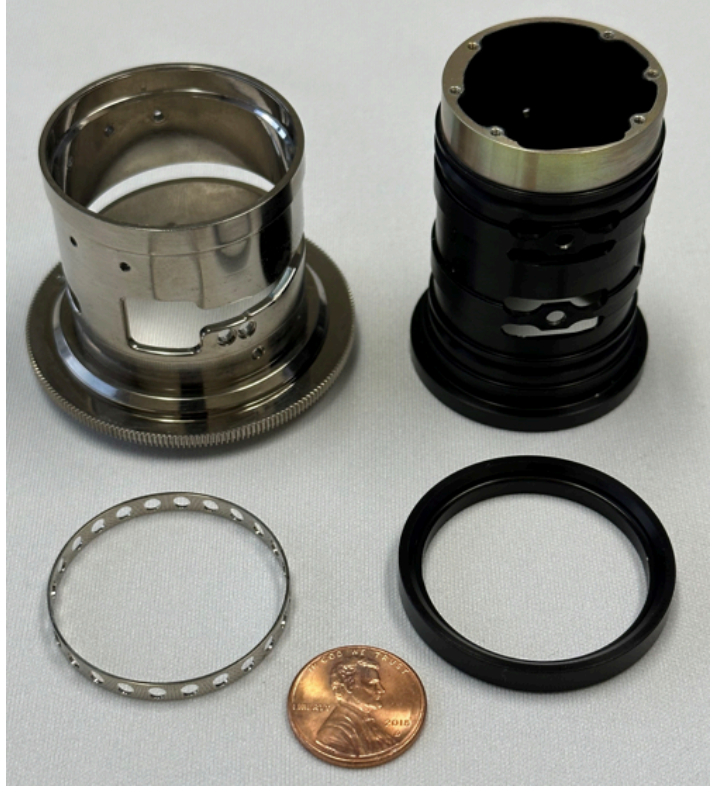

*Figure 9. Prototype hardware including (from left to right): bearing race ring, ball separator, cam barrel, inner optical structure.*

The cam barrel has a slot that is machined with a ramp profile specific to the optical system allowing the focus group to travel an appropriate distance for the object space distances it is trying to focus. A follower pin with an 1/8" diameter bearing is attached to the focus group and rides the surface of the cam profile such that when the cam barrel is rotated, the follower pin translates fore/aft. A custom compression spring preloads the bearing of the cam follower pin against one surface of the cam profile such that it is always bearing on the same side of the profile, except at the beginning of the profile where the launch lock exists. The launch lock engages the pin itself and removes the bearing from the load path, constraining the focus group during launch and rotorcraft flight.

The focus group is precision aligned to a linear way that rides a complementary precision-aligned linear rail. A snubber pin prevents excessive displacements of the flexured rail interface during launch loads. The radial position of the focus group is shimmed to the linear way, whereas tilt is managed through mechanical tolerances. Fixed group alignment is managed through mechanical tolerance alone. The linear rail is aligned to the inner optical structure using custom tooling through a flexured interface to maintain optical alignment and reduce stress-induced issues over the expected temperature range.

*Table 2. Tolerance requirements for optical groups within the focus mechanism*

| Group Alignment Tolerances | | |
|---|---|---|
| A1 | Fixed Group Axial Position | ±0.025 mm (±0.0010 in) |
| A2 | Fixed Goup Centration X, Y | ±0.020 mm (±0.0008 in) |
| A3 | Fixed Group Tilt X, Y | ±0.049 deg |
| A4 | Focus Group Centration X, Y | ±0.020 mm (±0.0008 in) |
| A5 | Focus Group Tilt X, Y | ±0.070 deg |
| A6 | Lens Filter Axial Position | ±0.250 mm (±0.0098 in) |
| A7 | Lens Filter Centration X, Y | ±0.254 mm (±0.010 in) |
| A8 | Lens Filter Tilt X, Y | ± 0.1 deg |
| A9 | Focus Group Range (E6-E7 airspace) | 15.897 - 21.26 mm |
| A10 | Focus Group Resolution (high) | 0.025 mm (.0010 in) |

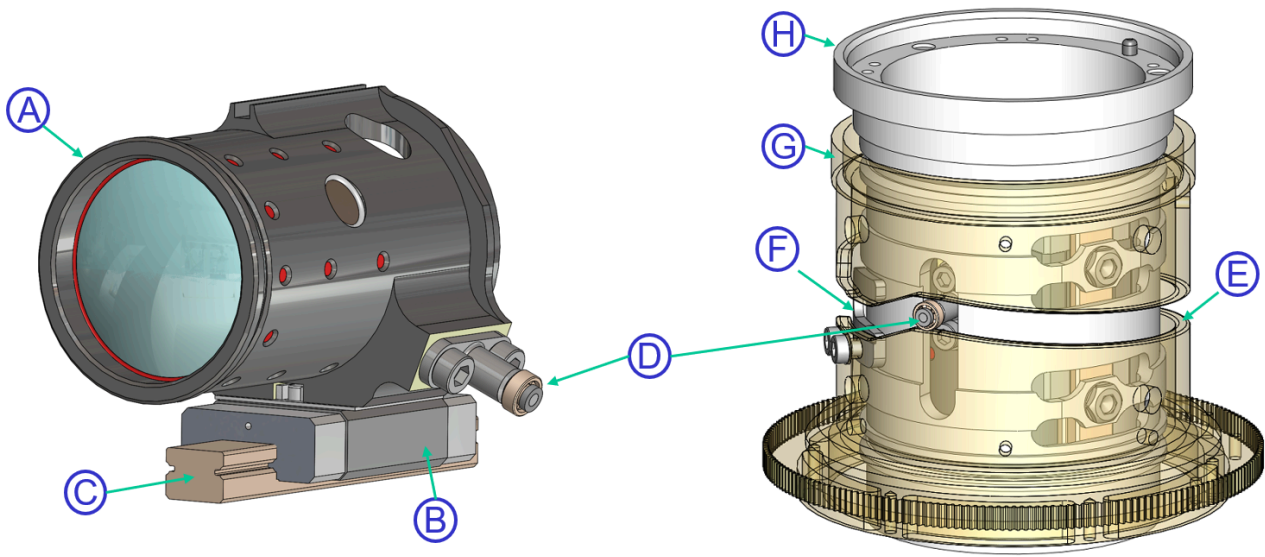

*Figure 10. Details of the Inner Optical Assembly of the Microscopic Camera Focus Mechanism including (a) moving group, (b) linear way, (c) linear rail, (d) cam follower pin, (e) cam profile, (f) launch lock, (g) cam barrel, (h) inner optical structure*

For position knowledge, a Hall effect sensor is mounted to the stationary exterior of the focus mechanism, while magnet pairs located at both ends of travel on the Cam barrel. The Hall sensor bracket provides adjustability of the sensor trigger point relative to the mechanism rotational position with the final position being set during assembly. Magnet pairs (as opposed to single magnets) to concentrate the trigger point location by focusing magnetic fields. Separate hard-stop components are incorporated that engage in the event of a Hall sensor failure to prevent damage to the mechanism. The hard stops are capable of supporting worst case motor stall load.

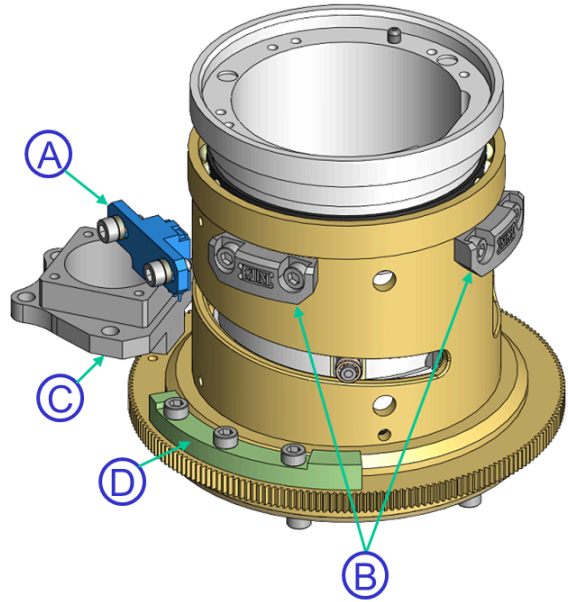

*Figure 11. Diagram showing positional knowledge components including (a) Hall effect sensor, (b) magnet pairs, (c) hard stop on motor mount, (d) hard stop on cam barrel*

**Optical**

The Microscopic Camera is required to image the Titan surface at high resolution from a distance of about a meter. This requires a relatively long focal length optical design focused close enough that the depth of field is only ±4mm, according to the hyperfocal distance equation. But the camera is required to image ±65mm, which correlates to surface variations of ±40mm. Some spacecraft instruments, such as the MER Microscopic Imager [13], have accomplished this by mounting the camera on a robotic arm and moving its position to focus on the subject. However, the Dragonfly spacecraft does not have a robotic arm. The camera is mounting to the body of the spacecraft, which drives the need for a focus mechanism. An additional twist, the camera is mounted at a 52 degree angle to the nominal Titan surface.

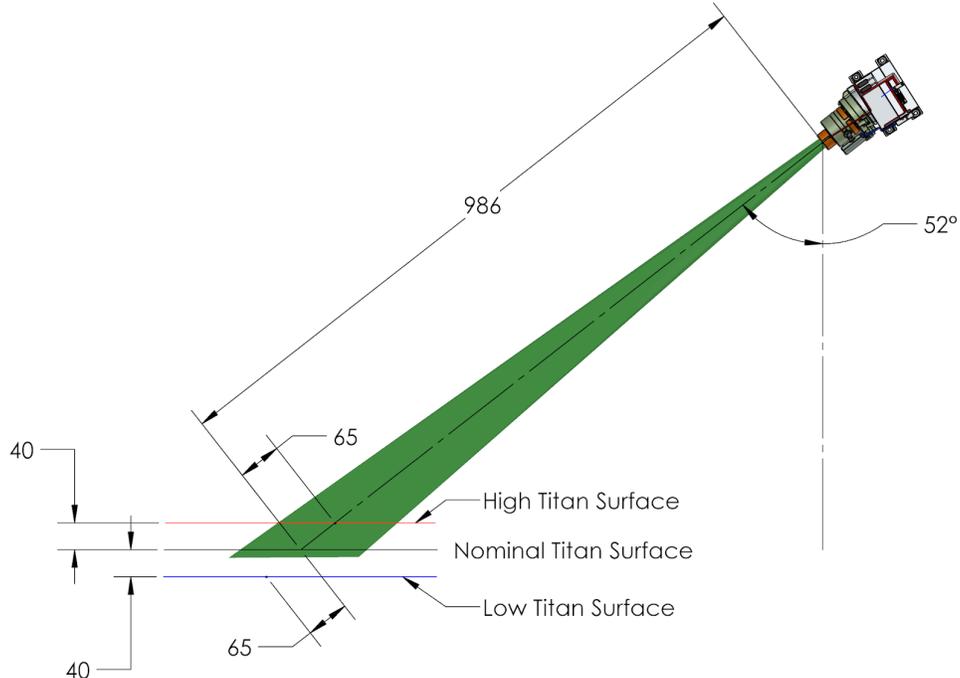

*Figure 12. Diagram illustrating the mounting constraints of the microscopic camera.*

Given the mounting constraints associated with the spacecraft and the desired performance, it was clear that the optical system would need a focus mechanism to be able to accommodate surface variations relative to its fixed mounting location. A 77.5mm design was developed with a 6-element fixed group and 3-element movable focus group. However, because the scene is at a relatively steep angle compared to the focal plane, a Scheimpflug optical design was incorporated to extend the depth of field. The incorporation of the Scheimpflug principle is enabled by tilting the boresight of the optics relative to the focal plane. The required tilt is machined into the electronics housing.

The optical system consists of 9 powered elements, including two 10$^{th}$ order aspheric surfaces. A fixed internal aperture stop is located in the fixed group yields an F/number of 5.6. The plano-plano fused silica window that closes out the mechanism also serves as the substrate for the 440-975mm bandpass filter. The 77.5mm focal length yields a nominal pixel scale of 60μm and a field of view of approximately 6 x 12 degrees.

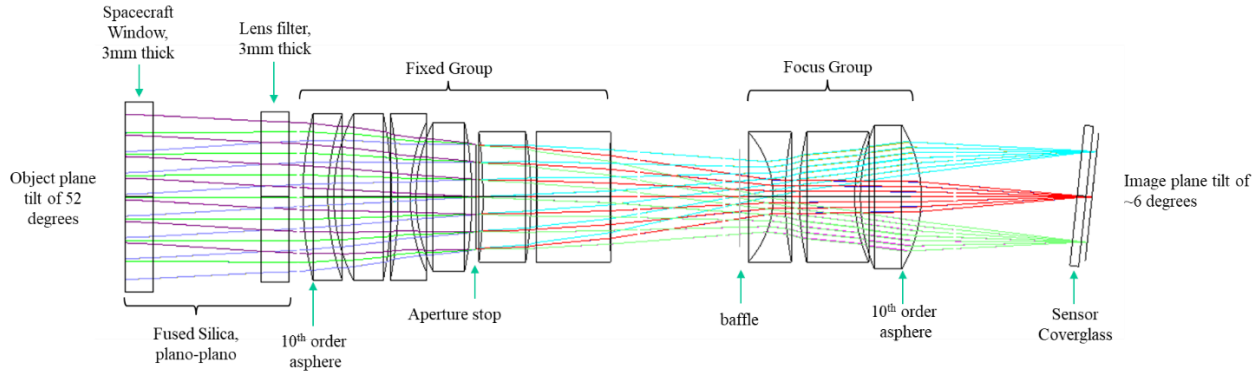

*Figure 13. Ray trace diagram of the Microscopic Camera optical groups, spacecraft window, mechanism window, and sensor.*

The design produces excellent MTF across the field. For the broadband imaging case, average MTF at the Nyquist frequency of the detector (104.2 lp/mm) is greater than 40%, before tolerancing. The thru-focus curve shows the performance is well balanced across the field. Note that the plots shown in Figure 14 are optics MTF, not system level MTF.

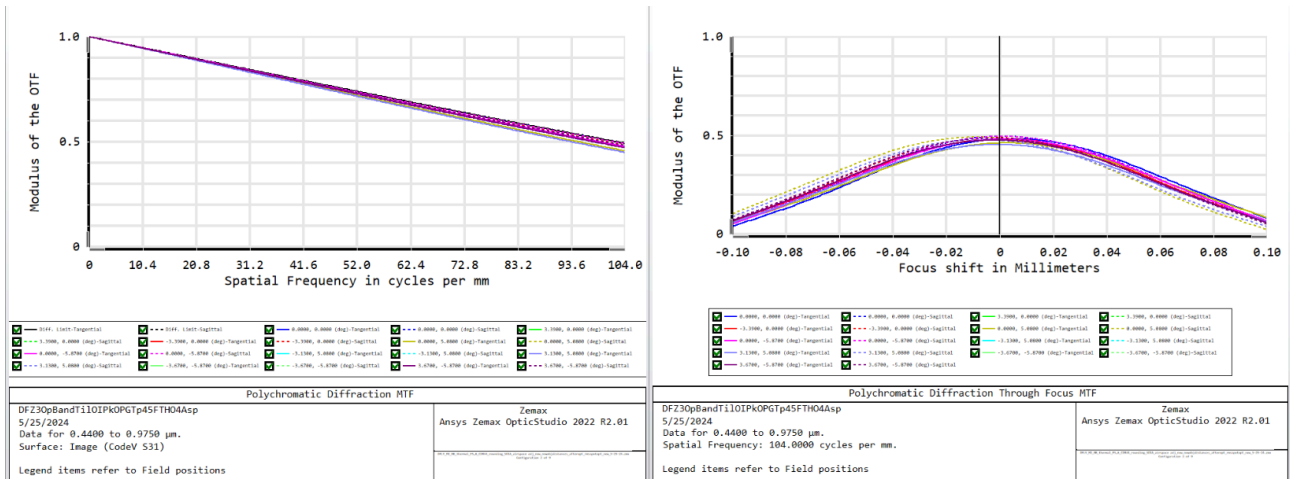

*Figure 14. Left: MTF versus Spatial Frequency, Right: Thru-focus plot of MTF at Nyquist at 20C and 986mm object distance, 440-975nm.*

For the night-time imaging case where the camera is imaging at narrower bandpasses using LED illumination, the optics MTF at Nyquist frequency exceeds the desired performance, even for the 935nm band. The MTF versus Spatial Frequency plot in Figure 15 shows the MTF for all field points for each LED band, all above 30% at the Nyquist frequency of the detector.

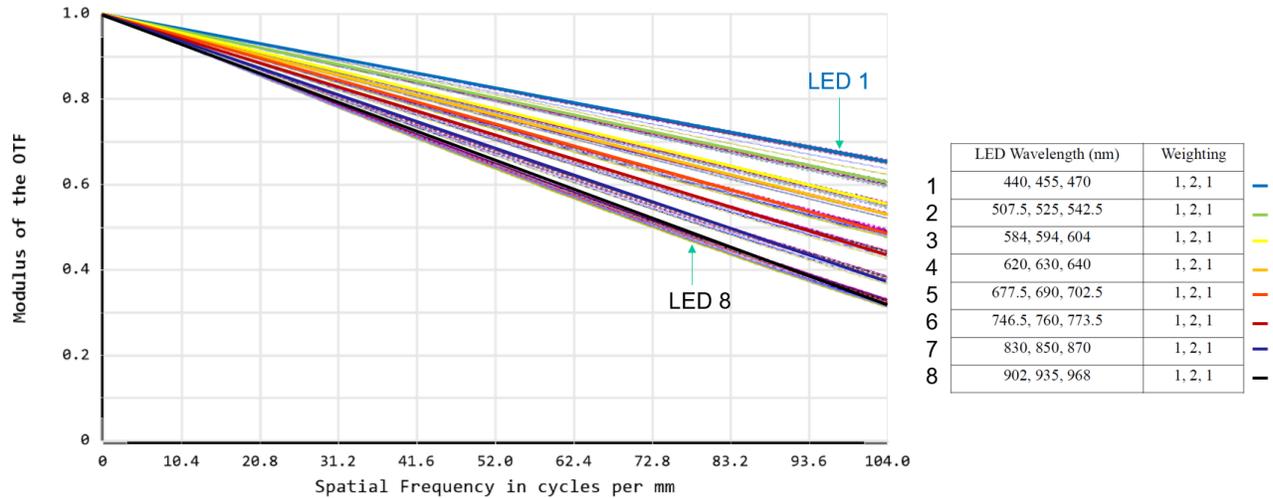

*Figure 15. MTF versus Spatial Frequency for each LED band at all field points.*

The average transmission of the lens is shown in Figure 16. The plot incorporates the transmission of the long-pass filter with a cut-on at 430nm and an out of band transmission of less than 1%. This blocks unwanted shortwave light from degrading MTF performance, but also ensures that the camera blocks the UV LED illumination, which has a 365nm wavelength. Additionally, the transmission analysis includes measured data from representative samples of extended broadband AR coatings. The resulting average transmission is greater than 85%.

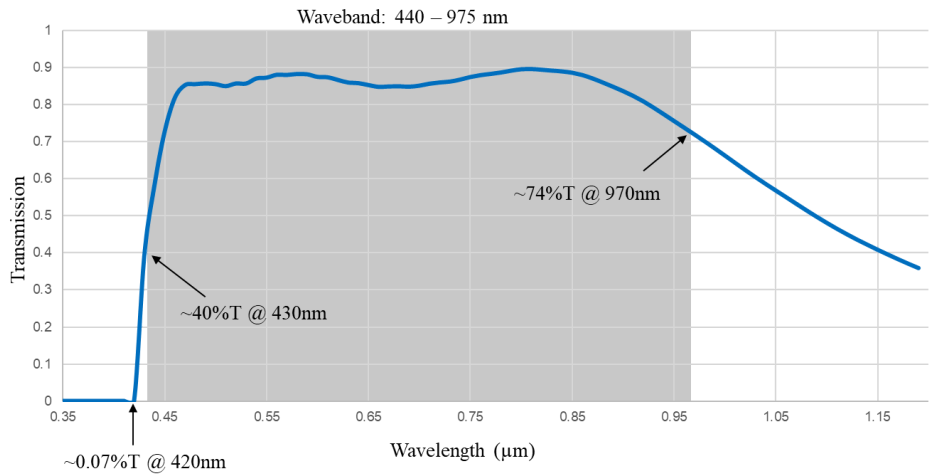

*Figure 16. Transmission versus Wavelength plot including as-built AR coatings and predicted long pass filter transmission.*

Because the camera is imaging a slanted surfaces (with respect to the focal plane), a keystone distortion effect is introduced. This causes the square image space coordinates to be transformed into a trapezoidal footprint in object space. This field of view meets the 50 x 50mm imaging footprint requirement with plenty of margin, but makes describing the field of view less straightforward than a typical optical system.

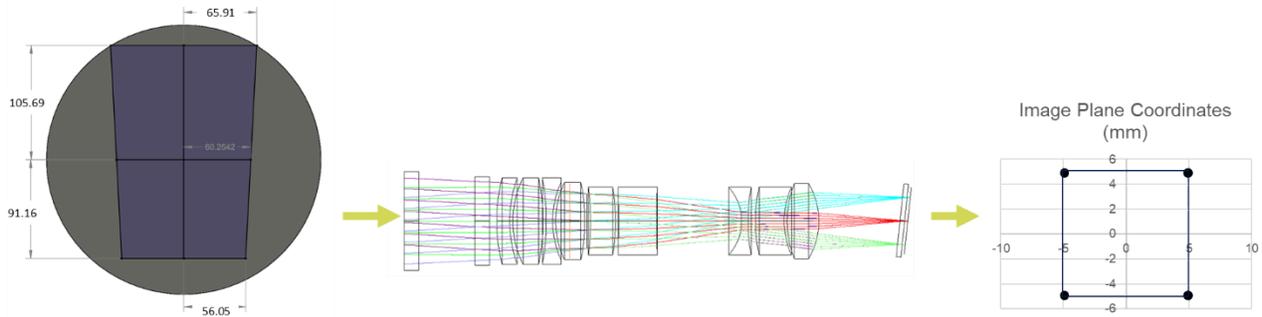

*Figure 17. Illustration of the image plane format (square box, right) and how it is transformed by the lens in object space to a keystone shaped footprint (purple trapezoid, left).*

A stray light analysis was performed with point sources and extended sources at different locations inside and outside of the field of view. Typical stray light mitigation techniques are employed to suppress unwanted ghost reflections within the lens. The interior of the lens is black anodized where possible. The aperture stop is effective in preventing the first four powered elements from being seen by the detector. Additionally, a baffle is located on the front of the focus group to prevent the seventh powered element from being illuminated by unwanted raypaths. Extended broadband AR coatings are used on all elements to minimize reflections. An example of the effectiveness of these measures is shown in Figure 18. This particular example uses an off axis point source and illustrates the ghost reflections that result from it. The figure is logarithmic scale to visualize the ghost shapes. All ghost reflections encountered in the stray light analysis were over three orders of magnitude lower than the primary image irradiance, thus not polluting the intended scene.

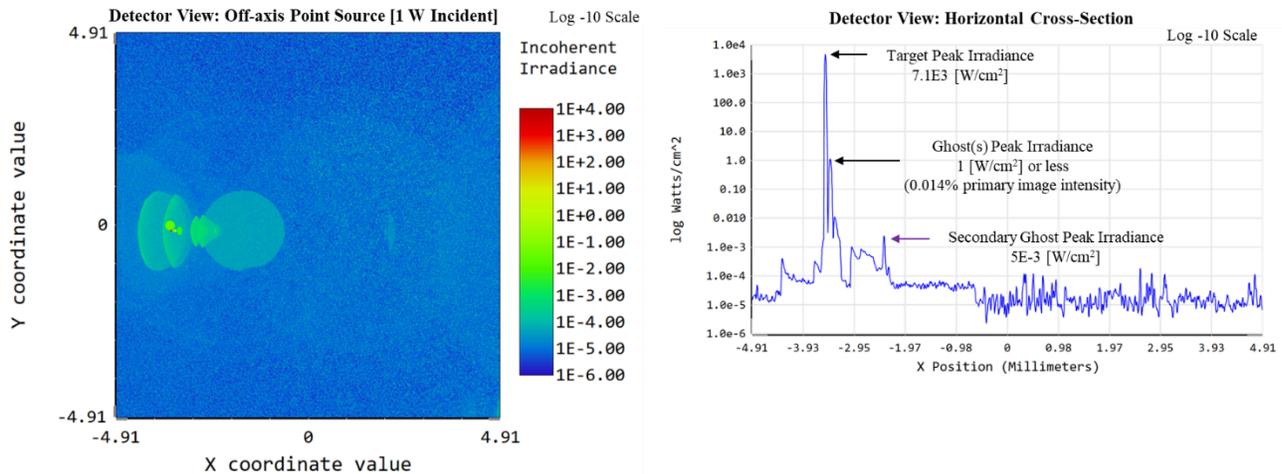

*Figure 18. One (of many) stray light analysis setups identifying ghost reflections from an off-axis point source*

Early in the program, a risk was identified regarding thermal gradients on the optics, particularly given that some optical systems that MSSS has flown in the past have been incredibly sensitive to gradients [14]. When the electronics are warmed up to operating temperature, the lens is still millimeters away from the -179°C Titan environment, shielded by a 3mm thick piece of glass. This causes substantial thermal gradients on the optics compared to previous programs. Out of the nine cameras on Dragonfly, the Microscopic Camera has the least severe gradients due to its small window size, the additional mechanism window between the spacecraft and the powered elements, and the fact that it has a focus mechanism that can focus out focal shifts. A Structural-Thermal-Optical-Performance (STOP) analysis was performed by taking the thermal gradients from the thermal model and applying them to the structural model. Distortions and/or rigid body motion of lens elements were exported and applied to the optical model. The lens elements are modeled as Gradient Index (GRIN) materials to take into account the change in index with respect to temperature and the optical performance is calculated to assess performance impacts at a given temperature. This data flow is visualized in Figure 19.

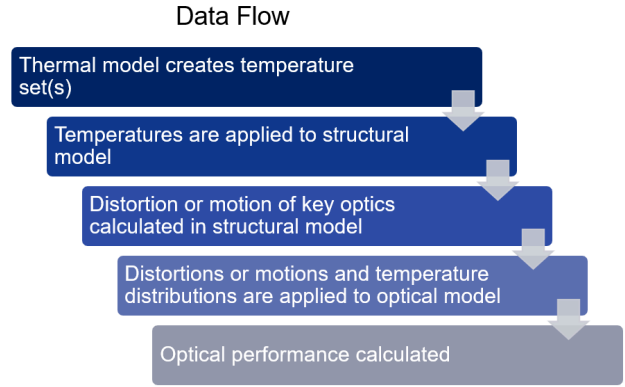

*Figure 19. Data flow as it pertains to performing STOP analysis of the Microscopic Camera*

A detailed thermal model was created that incorporated the spacecraft enclosure, window, and notional insulation, along with the Microscopic Camera. Convection of the Titan environment was modeled on the window and insulation of the spacecraft for two cases: flight through the Titan atmosphere and landed wind flow. Natural convection inside the enclosure cavity was included between the spacecraft window, spacecraft enclosure, and the camera. Gas conduction was modeled in small gaps where convective currents cannot form, such as the airspace between powered elements 1 and 2. Radiation was included inside the electronics and lens barrel, as well as on external surfaces facing the Titan environment. The camera was thermal isolated from the spacecraft using a conductance value that kept the camera at -85°C during hibernation. Finally, the heater power necessary to warm the electronics to >-20°C within 60 minutes was determined.

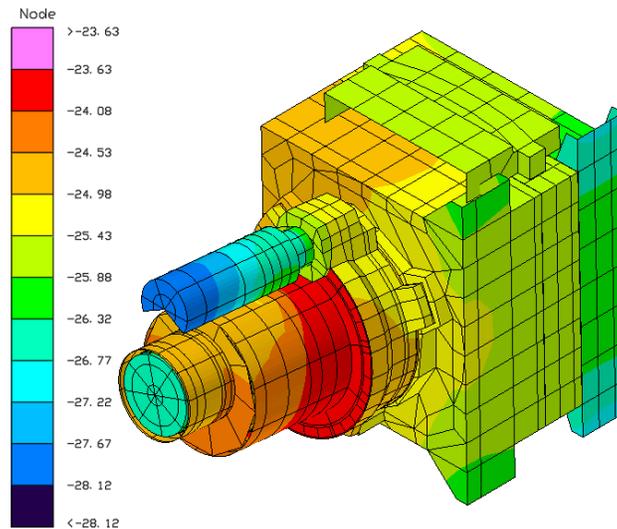

*Figure 20. Temperature plot of the Microscopic Camera after 60 minute warmup to operating temperature.*

The plano-plano mechanism window exhibits a 1.3°C peak gradient after the 60 minute warmup. The effect of this is relatively benign given that the element is unpowered and shields the powered elements behind it, which is not the case with the other DragonCam cameras whose powered front elements are right behind the spacecraft window. The resulting thermal gradients for individual powered elements are <0.5°C, whereas the axial gradient between all the powered elements is about 4°C as can be seen in Figure 21.

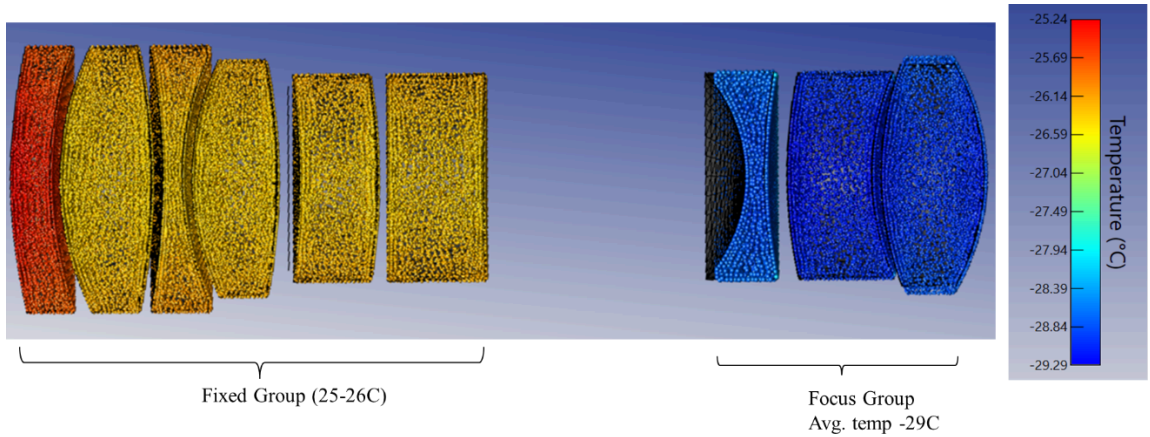

*Figure 21. Temperature plot of the powered elements inside the Microscopic Camera focus mechanism.*

The temperatures from the thermal model are mapped onto the structural model, which can calculate displacements and deformations of the optical elements and housings. The displacements/deformations are applied to the optical model and the temperature data is applied to the optical model as a point cloud, allowing the index of the material to be modeled as a function of temperature within the optical model. When the performance is calculated with these effects incorporated, the average MTF is impacted by about 2%. This is after the unit is refocused by -140μm to recover the focus shift.

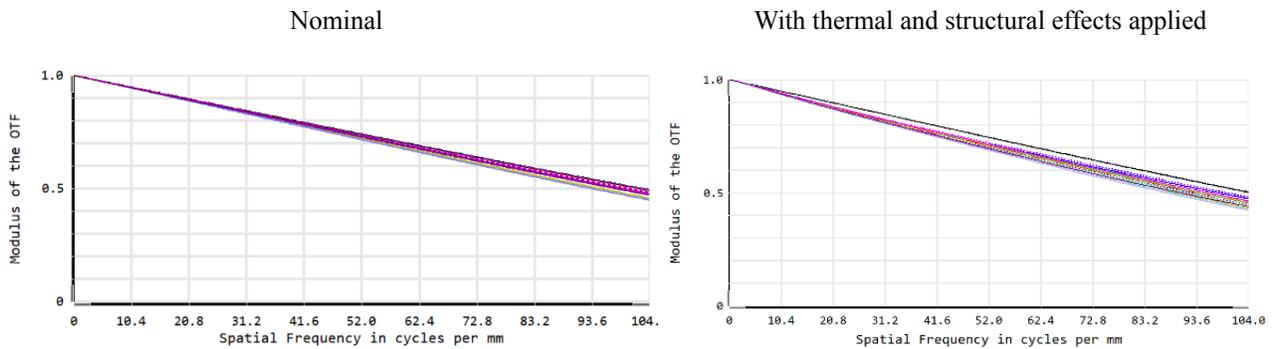

*Figure 22. MTF versus Spatial Frequency before temperature and structural effects are applied (left) and after (right).*

Prototype lens builds are shown in Figure 23 with a penny for scale.

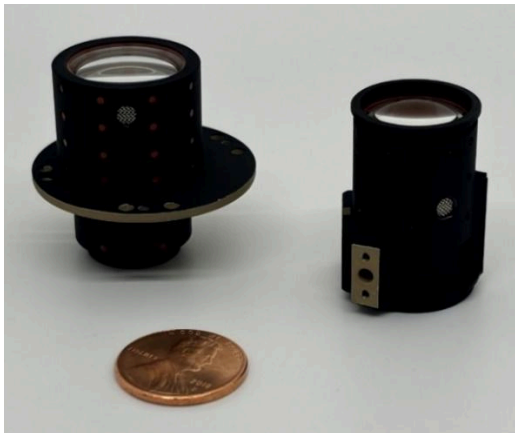
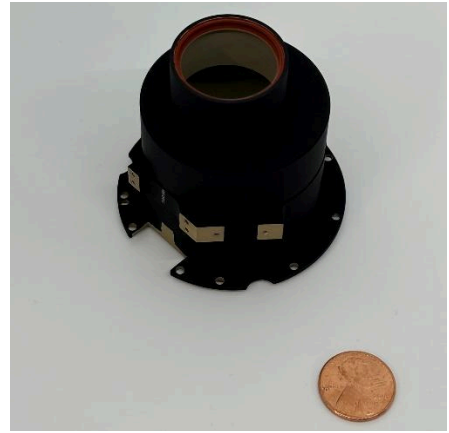

*Figure 23. Prototype hardware including the assembled fixed and moving group (left) and mechanism top housing with bandpass filter (right)*

**Electrical**

The ECAM-L50 electronics is built from three printed circuit assemblies: the Interface Adaptor (L50IFA), Digital Module (L50DM), and Focal Plane Assembly (L50FPA), as shown in Figure 1. The L50IFA hosts an ECAM SpaceWire interface with an optional second interface for increased throughput and redundancy. The IFA also includes a temperature sensor located near the Micro-D connector, with good thermal conductivity to the enclosure stack. On the IFA, the 5V power input is regulated to +3.3 V, +2.5 V, +1.8V and +1.5 V, then passed on to the Digital Module. The L50IFA2 provides drive to an off-board three-phase motor, under control of the FPGA, in addition to the functionality described above.

The primary purpose of the L50DM is to host the FPGA. Two oscillators drive the system clocks of the FPGA. A supervisory IC provides a power-on reset timer that is input to the FPGA as the master reset, ensuring the clock oscillator has fully stabilized before the logic is released from reset. The DM also accommodates four NAND flash banks.

The L50FPA contains three Line Current Limiter (LCL) modules that control the three power supplies to the sensor. The LCLs serve a primary function of controlling power switching so that the three supplies can be appropriately sequenced by FPGA control and a secondary function of monitoring supply current and quickly dropping the supply voltage in the event that a preset threshold is exceeded. This serves to mitigate the risk of damage due to any off-nominal operating case that would cause excessive supply current. LCL protection status is reported to the FPGA so that the power supplies, clock, and reset signals can be sequenced off by FPGA logic when a protection event occurs on any one of the supplies. The host system can monitor LCL status and must intervene to power the sensor back on and re-initialize it after a protection event occurs.

Perhaps the most important component, the image sensor is also located on the L50FPA. The Microscopic Camera image sensor uses a custom packaged color CMOS sensor. While the same sensor is used in multiple other cameras (including the LCAM camera that performed Terrain Relative Navigation for the Mars 2020 Perseverance Rover [8]), the heritage Ceramic Leadless Chip Carrier package was concerning given the broad temperature range that the camera would be cycled through on the Titan surface.

The sensor used in the Dragonfly Microscopic Camera is the same as is used in MSSS's ECAM-P50 [5], LanderCam (LCAM) [16], and Docking Camera (DCAM) [17] Cameras. For all of these cameras, the commercial sensor die is procured as a wafer, diced, and packaged in a custom package. This allows for several customizations to improve the reliability and performance of the sensor. The primary reason for repackaging the sensor is to achieve a more uniform and controlled die attach. Secondly, a more robust, AR coated coverglass is able to be incorporated. And in the case of Dragonfly, the sensor die was repackaged in a Ceramic Quad Flat Pack that has a lead configuration less susceptible to thermal stress fatigue (and, of less importance in this application, vibration fatigue). However, the Quad Flat Pack is a lead formed part and introduces alignment error in the optical stackup. To control this, custom-designed fixturing with 6-degree-of-freedom alignment capability was developed to place the sensor die parallel to the PCB using a depth-measuring microscope. The PCB is controlled to the housing with mechanical tolerances.

**Z-stacking**

The Microscopic Camera has a limited depth of field. For example, a 77.5mm lens focused at 1m has about a ±4mm depth of field when calculated using the hyperfocal distance equation [15]. The Scheimpflug principle is utilized to offset the tilted object plane, allowing the entire image plane to be in focus instead of a horizontal sliver. However, there may be imaging scenarios where the depth of field is not adequate. For these scenarios, Z-stacking (or focus merging) will be used to image the scene, where multiple images are taken at different focus distances to compile a composite image and the final image is compiled in software.

As opposed to being mounted on a robotic arm where the camera distance-to-object can be changed, the Microscopic Camera is fixed-mounted to the side of the Dragonfly spacecraft, so the only method of changing focus is within the camera. A study was performed over the nominal 921mm to 1051mm object distance range to determine the number of images needed to form a composite image. The depth of field extents for each focus distance was defined where the

MTF falls to half the best-focus MTF. Using this criteria, the DOF is calculated to be 17.5mm and thus 9 images are required to cover the 921 – 1051mm object distance range. If peak MTF is desired across the field, where no noticeable dropoff in MTF occurs across the depth of field or across the field of the detector, the number of images increases to 18.

To assemble the focus stack, individual images can be acquired and relayed to Earth, or the focus merge can be performed inside the DVR, then the composite image can be downlinked. The latter is anticipated to be preferred for Dragonfly as the spacecraft is downlink constrained and the focus merging serves as a form of data compression. To acquire the focus stack, the instrument is commanded to acquire images at each focus position and stores the raw images to flash memory. This is accomplished by specifying a starting motor position for the lens moving group, the number of images to acquire, and the number of motor counts to step through between each image. Once the images have been acquired, the DVR can be commanded to perform the focus merge process. First, Bayer color interpolation is performed on each image. Then each image is split into $Y:C_R:C_B$ components and the focus merge is performed with the $Y:C_R:C_B$ data using a windowed Sum-Modified-Laplacian focus measure to determine the areas of best focus [18].

## ACKNOWLEDGEMENTS

Malin Space Science Systems would like to acknowledge NASA for funding this ambitious mission, APL for envisioning and carrying out the mission, Motiv Space Systems for the design and assembly of the focus mechanism, Collins Aerospace for the design and assembly of the optics, and Synopsis for preliminary optical design and review of the flight design.